\documentclass[12pt]{article}
\begin{document}

\begin{center}{\Large\bf New Inequalities for Tomograms in the
Probability  Representation of Quantum States}
\end{center}
\begin{center}{{\bf Sergio De Nicola},$^1$ {\bf Renato Fedele},$^2$
{\bf Margarita A. Man'ko}$^3$ and {\bf Vladimir I. Man'ko}$^3$}
\end{center}
\begin{center}{ \it ${}^1$ Istituto di Cibernetica ``Eduardo Caianiello'' del
CNR Comprensorio ``A.~Olivetti'' Fabbr.~70, Via Campi Flegrei~34,
I-80078 Pozzuoli (NA), Italy\\ ${}^2$ Dipartimento di Scienze
Fisiche, Universit\`{a} ``Federico II'' di Napoli\\ and Istituto
Nazionale di Fisica Nucleare, Sezione di Napoli
\\ Complesso Universitario di M. S. Angelo, Via Cintia, I-80126 Napoli,
Italy\\ ${}^3$ P. N. Lebedev Physical Institute\\ Leninskii
Prospect 53, Moscow 119991 Russia}
\end{center}
Emails:

s.denicola@cib.na.cnr.it

renato.fedele@na.infn.it

mmanko@sci.lebedev.ru

manko@sci.lebedev.ru

\begin{abstract}
New inequalities for symplectic  tomograms of quantum states and
their connection with entropic uncertainty relations are discussed
within the framework of the probability representation of quantum
mechanics.
\end{abstract}

{\bf keywords}: tomographic map,  symplectic tomography, entropic
uncertainty relations, inequalities for tomograms.

\section{Introduction}

There exists several equivalent formulations of quantum mechanics
(see, for example, review~\cite{AmJP}). Recently a new formulation
of quantum mechanics, which is called the probability representation
of quantum mechanics, was introduced~\cite{PLA96,FP97}. Within the
framework of this formulation, the quantum states are described by
standard probability distributions instead of wave functions or
density matrices. The probability representation of quantum
mechanics is equivalent to all other representations but it is more
convenient for considering some class of quantum problems where one
can use the well-elaborated mathematical tools of the probability
theory.

The Shannon entropy~\cite{Shannon} is the functional characteristics
of any probability distribution and it was used to introduce the
tomographic entropy for continuous variable~\cite{OlVolJRLR97} and
for analytic signals~\cite{Rita-entropy}, as well as for spin
tomographic probabilities~\cite{Olga-spin}. The probability
representation was shown~\cite{Marmo-PhysScr,Marmo-JPA,Marmo-Krakow}
to realize a new version of the quantization procedure based on the
star-product formalism. The star-product approach and different
properties of tomographic entropies within the quantum-information
framework were studied in \cite{Rui-JRLR} where
R\`{e}nyi~\cite{renyi} entropy was considered using the
spin-tomographic probabilities (spin tomograms).

There exist inequalities for Shannon entropies associated with
probability distributions of the position and momentum (see, for
example, review~\cite{183}). Recently~\cite{Balq-ph} new entropic
uncertainty relations based on properties of R\`{e}nyi~\cite{renyi}
entropy were found. The entropic uncertainty relations were used to
obtain new inequalities for tomographic probability distributions
for continuous variables (called symplectic
tomograms)~\cite{EJPB06}.

The aim of this work is to give a review of new entropic uncertainty
relations and to obtain new inequalities for tomograms of quantum
states for the case of several modes.

The paper is organized as follows.

In Section~2 we discuss the known entropic uncertainty relations for
the one-mode and multimode cases. In Section~3 we review the
properties of tomographic
entropies~\cite{OlVolJRLR97,Rita-entropy,Olga-spin}. In Section~4 we
consider the integral inequalities for symplectic tomograms in the
one-mode case. In Section~5 we study the  the integral inequalities
for both optical and symplectic tomograms of multimode quantum
states. The conclusions are presented in Section~6.

\section{Entropic Uncertainty Relations}

If the quantum state of a particle is described by a wave function
$\psi(x)$ in the position representation (or a wave function
$\widetilde\psi(p)$ in the momentum representation), the Shannon
entropies $S_x$ and $S_p$ connected with two probability
distributions $|\psi(x)|^2$ and $|\widetilde\psi(p)|^2$ are given by
the integrals:
\begin{eqnarray}
S_x&=&-\int|\psi(x)|^2\ln|\psi(x)|^2,\nonumber\\
&&\label{1}\\
S_p&=&-\int|\widetilde\psi(p)|^2\ln|\widetilde\psi(p)|^2.\nonumber
\end{eqnarray}
The entropies satisfy the entropic uncertainty
relation~\cite{Hirsch57,Bal-Bir75,183}:
\begin{equation}\label{2}
S_x+S_p\geq \ln\pi e.
\end{equation}
In the case of density matrices $\rho(x,y)$ and $\rho(p_x,p_y)$ of
the quantum state (given in the position and momentum
representations, respectively), the Shannon entropies are defined as
follows:
\begin{eqnarray}
S_x&=&-\int\rho(x,x)\ln\rho(x,x)\,dx,\nonumber\\
&&\label{3}\\
S_p&=&-\int\rho(p,p)\ln\rho(p,p)\,dp. \nonumber
\end{eqnarray}
These entropies satisfy the same inequality~(\ref{2}).

In the case of multimode states, the entropies
\begin{eqnarray}
S_{\vec x}&=&-\int\rho(\vec x,\vec x)\ln\rho(\vec x,\vec x)\,d{\vec
x},
\nonumber\\
&&\label{4}\\
S_{\vec p}&=&-\int\rho(\vec p,\vec p)\ln\rho(\vec p,\vec p)\,d{\vec
p} \nonumber
\end{eqnarray}
satisfy the entropic uncertainty relation with extra factor
\begin{equation}\label{5}
S_{\vec x}+S_{\vec p}\geq N\ln\pi e,
\end{equation}
where $N$ is the number of modes.

In fact, the entropic uncertainty relations (\ref{2}) and (\ref{5})
can be interpreted as constrains for density matrices. These
constrains are connected with the positivity conditions of the
density operator of any quantum state.

\section{Tomograms and Tomographic Entropies}

In \cite{PLA96} the new formulation of quantum mechanics was
suggested. Within the framework of this formulation, the quantum
state described by the tomographic-probability distribution
$w(X,\mu,\nu)$ (called symplectic tomogram) relates to a density
operator $\hat\rho$ by the formula~\cite{RitaJPA}
\begin{equation}\label{6}
w(X,\mu,\nu)=\mbox{Tr}\,\hat\rho~\delta(X-\mu\hat q-\nu\hat p).
\end{equation}
The inverse transform reads~\cite{Dariano96}
\begin{equation}\label{7}
\hat\rho=\frac{1}{2\pi}\int w(X,\mu,\nu)\exp\left[i(X-\mu\hat
q-\nu\hat p)\right]\,dX\,d\mu\,d\nu,
\end{equation}
where $\hat q$ and $\hat p$ are the position and momentum operators,
respectively, and $X$, $\mu$, and $\nu$ are real variables. The
variable $X$ is a random position measured in a reference frame in
the phase space labeled by two real parameters $\mu=s\cos\theta$ and
$\nu=s^{-1}\sin\theta$. The angle $\theta$ is the rotation angle of
the axis in the phase space and the scaling parameter $s$ determines
a new scale in the reference frame. Thus, one has the nonnegativity
condition
\begin{equation}\label{8}
w(X,\mu,\nu)\geq 0\end{equation} and the normalization condition of
the tomographic-probability density
\begin{equation}\label{9}
\int w(X,\mu,\nu)\,dX=1.\end{equation}

If $s=1$, the tomogram is called optical tomogram and it is used for
measuring the quantum states of photons~\cite{Raymer}
\begin{equation}\label{10}
w(X,\theta)=\mbox{Tr}\,\hat\rho~\delta\left(X-\hat q\cos\theta -\hat
p\sin\theta\right).
\end{equation}
One has
\begin{equation}\label{11}
w(X,\mu=\cos\theta,\nu=\sin\theta)=w(X,\theta).\end{equation} The
symplectic tomogram satisfies the homogeneity condition
\begin{equation}\label{12}
w(\lambda
X,\lambda\mu,\lambda\nu)=\frac{1}{|\lambda|}w(X,\mu,\nu).\end{equation}

Since the optical tomogram $w(X,\theta)$ and symplectic tomogram
$w(X,\mu,\nu)$ are standard probability densities, the Shannon
definition was used in \cite{OlVolJRLR97,Rita-entropy} to introduce
the tomographic entropies
\begin{equation}\label{13}
S(\mu,\nu)=-\int w(X,\mu,\nu)\,\ln w(X,\mu,\nu)\,dX
\end{equation}
and
\begin{equation}\label{14}
S(\theta)=-\int w(X,\theta)\,\ln w(X,\theta)\,dX.
\end{equation}
The von Neuman entropy of quantum state
\begin{equation}\label{15}
S_N=-\mbox{Tr}\,\hat\rho~\ln\hat\rho
\end{equation}
is equal to zero for all pure quantum states.

The tomographic entropies $S(\theta)$ and $S(\mu,\nu)$ distinguish
different pure states.

The homogeneity property of tomogram (\ref{12}) yields the following
property of tomographic entropy~\cite{ActaHung06}
\begin{equation}\label{16}
S\Big(\sqrt{\mu^2+\nu^2}\cos\theta,\sqrt{\mu^2+\nu^2}\sin\theta\Big)
-\frac{1}{2}\,\ln(\mu^2+\nu^2)=f(\theta).
\end{equation}
This means that effectively the tomographic entropy depends on angle
variable only.

\section{Tomographic Entropic Uncertainty Relation for One Mode}

Recently~\cite{EJPB06,ActaHung06} new inequalities were obtained for
tomographic entropies and tomograms of quantum states for continuous
variables. We present here these inequalities for the one-mode case.
As it was shown in \cite{Rita-entropy} the tomogram of quantum state
(\ref{6}) can be considered as the probability distribution of
position for the state of ``artificial quantum harmonic oscillator''
evolving from some initial state $\hat\rho(0)$ to the state
$\hat\rho(t)$. In view of this observation, the periodic-in-time
motion of the oscillator provides the change of the position
probability density into the momentum probability density after
evolving one quarter of the vibration period. Thus, the entropies
and their inequalities (\ref{2}) can be calculated for tomograms of
any quantum state providing the following inequality relation:
\begin{equation}\label{17}
S(\theta)+S(\theta+\pi/2)\geq \ln\pi e.
\end{equation}
This inequality means the integral condition for optical tomogram of
quantum state
\begin{equation}\label{18}
\int\left[w(X,\theta)\ln w(X,\theta)+w(X,\theta+\pi/2) \ln
w(X,\theta+\pi/2)\right]\,dX+\ln\pi e\leq 0.
\end{equation}
The optical tomogram was measured in the experiments with
photons~\cite{Raymer} and now inequality (\ref{17}) can be used for
extra check of the experimental data obtained.

\section{Inequalities with Extra Parameters for Tomograms}

In \cite{Balq-ph} the new uncertainty relation was obtained for
R\`{e}nyi entropy related to the probability distributions for
position and momentum of quantum state with density operator
$\hat\rho$. The uncertainty relation reads
\begin{eqnarray}\label{B1}
&&\frac{1}{1-\alpha}\,\ln\left(\int_{-\infty}^\infty
dp\left[\rho(p,p)\right]^\alpha\right)
+\frac{1}{1-\beta}\,\ln\left(\int_{-\infty}^\infty
dx\left[\rho(x,x)\right]^\beta\right)\nonumber\\
&&\geq -\frac{1}{2(1-\alpha)}\,\ln\frac{\alpha}{\pi}
-\frac{1}{2(1-\beta)}\,\ln\frac{\beta}{\pi}\,,\end{eqnarray} where
positive parameters $\alpha$ and $\beta$ satisfy the
constrain
\begin{equation}\label{B2}
\frac{1}{\alpha}+\frac{1}{\beta}=2.
\end{equation}
Using the same argument that we employed to obtain inequality
(\ref{18}), one arrived at the condition for optical
tomogram~\cite{ActaHung06}
\begin{eqnarray}\label{B4}
&&\frac{q-1}{q}\,\ln\left\{\int_{-\infty}^\infty
dX\left[w\left(X,\theta+\frac{\pi}{2}\right)\right]^{1/(1-q)}\right\}\nonumber\\
&&+ \frac{q +1}{q}\,\ln\left\{\int_{-\infty}^\infty
dX\left[w\left(X,\theta\right)\right]^{1/(1+q)}\right\}\nonumber\\
&&\geq\frac{1}{2}\left\{\frac{q-1}{q}\,\ln\,[\pi(1-q)]+\frac{q+1}{q}\,\ln\,
[\pi(1+q)]\right\},\end{eqnarray} where the parameter $q$ is defined
by $\alpha=(1-q)^{-1}$.

Below we present a new inequality for symplectic tomogram in the
one-mode case. It reads
\begin{eqnarray}\label{B5}
&&\frac{q-1}{q}\,\ln\left\{\int_{-\infty}^\infty
dX\left[w\left(X,-\sqrt{\mu^2+\nu^2}\sin\theta,
\sqrt{\mu^2+\nu^2}\cos\theta\right)\right]^{1/(1-q)}\right\}\nonumber\\
&&+ \frac{q +1}{q}\,\ln\left\{\int_{-\infty}^\infty
dX\left[w\left(X,\sqrt{\mu^2+\nu^2}\cos\theta,
\sqrt{\mu^2+\nu^2}\sin\theta\right)\right]^{1/(1+q)}\right\}\nonumber\\
&&\geq\frac{1}{2}\left\{\frac{q-1}{q}\,\ln\,[\pi(1-q)]+\frac{q+1}{q}\,\ln\,
[\pi(1+q)]\right\}.\end{eqnarray} This inequality can be interpreted
as a generalization of the inequality (\ref{18}) extended from
optical tomogram to symplectic tomogram.

In view of the inequality for R\`{e}nyi entropy adopted from
\cite{Balq-ph}, the above condition for tomogram of quantum state
can be generalized for the multimode case as well. Then for
symplectic tomogram of quantum state with density operator
$\hat\rho$ defined as
\begin{eqnarray}\label{B6}
&&w\left(X_1,X_2,\ldots,X_N,\mu_1,\mu_2,\ldots,\mu_N,\nu_1,\nu_2,
\ldots,\nu_N\right)\nonumber\\
&&=\mbox{Tr}\,\left[\hat\rho\,\delta\left(X_1-\mu_1\hat
q_1-\nu_1\hat p_1\right)\,\delta\left(X_2-\mu_2\hat q_2-\nu_2\hat
p_2\right)\cdots\,\delta\left(X_N-\mu_N\hat q_N-\nu_N\hat
p_N\right)\right]\nonumber\\
&&\end{eqnarray} where $\hat q_k$ and $\hat p_k$ $(k=1,2,\ldots, N)$
are position and momentum operators, respectively, we obtain
\begin{eqnarray}\label{B7}
&&\frac{q-1}{q}\,\ln\left\{\int_{-\infty}^\infty d{\vec
X}\left[w\left(X_1,X_2,\ldots,X_N,-\sqrt{\mu_1^2+\nu_1^2}\sin\theta_1,\ldots,
\right.\right.\right.\nonumber\\
&&\left.\left.\left.-\sqrt{\mu_N^2+\nu_N^2}\sin\theta_N,\sqrt{\mu_1^2+\nu_1^2}\cos\theta_1,\ldots,
\sqrt{\mu_N^2+\nu_N^2}\cos\theta_N\right)\right]^{1/(1-q)}\right\}\nonumber\\
&&+\frac{q +1}{q}\,\ln\left\{\int_{-\infty}^\infty d{\vec
X}\left[w\left(X_1,X_2,\ldots,X_N,\sqrt{\mu_1^2+\nu_1^2}\cos\theta_1,
\ldots,\right.\right.\right.\nonumber\\
&&\left.\left.\left.\sqrt{\mu_N^2+\nu_N^2}\cos\theta_N,
\sqrt{\mu_1^2+\nu_1^2}\sin\theta_1,
\ldots,\sqrt{\mu_N^2+\nu_N^2}\sin\theta_N
\right)\right]^{1/(1+q)}\right\}\nonumber\\
&&\geq\frac{N}{2}\left\{\frac{q-1}{q}\,\ln\,[\pi(1-q)]+\frac{q+1}{q}\,\ln\,
[\pi(1+q)]\right\}.\end{eqnarray}

For optical tomogram
$w\left(X_1,\ldots,X_N,\theta_1,\ldots,\theta_N\right)$, the
inequality reads
\begin{eqnarray}\label{B8}
&&\frac{q-1}{q}\,\ln\left\{\int_{-\infty}^\infty d\vec
X\left[w\left(X_1,X_2,\ldots,X_N,\theta_1,\ldots,\theta_N\right)\right]^{1/(1-q)}
\right\}\nonumber\\
&&+\frac{q+1}{q}\,\ln\left\{\int_{-\infty}^\infty d\vec
X\left[w\left(X_1,X_2,\ldots,X_N,\theta_1+\pi/2,\ldots,\theta_N+\pi/2\right)\right]^{1/(1+q)}
\right\}\nonumber\\
&&\geq\frac{N}{2}\left\{\frac{q-1}{q}\,\ln\,[\pi(1-q)]+\frac{q+1}{q}\,\ln\,
[\pi(1+q)]\right\}.\end{eqnarray}

Inequalities (\ref{B7}) and (\ref{B8}) are saturated for Gaussian
tomograms. In the limit $q\to 0$, they become entropic uncertainty
relations found in \cite{EJPB06,ActaHung06}.

\section{Conclusions}

To conclude, we summarize the main results of our study.

The new uncertainty relations were reviewed within the framework of
the probability representation of quantum mechanics. The entropic
uncertainty relations have the form of integral condition for
tomograms of quantum states which contain the complete information
on the states. The new inequality containing extra parameter were
obtained for some integral expressions containing the quantum state
tomograms on the base of recently found~\cite{Balq-ph} uncertainty
relations for R\'enyi entropy of quantum states. The conditions for
the one-mode and multimode optical tomograms are of particular
interest since these tomograms are directly measured in
quantum-optics experiments~\cite{Raymer}. We hope to get analogous
new inequalities for tomograms depending on discrete variables.

\section*{Acknowledgments}

M.A.M. thanks the Organizers of the International Workshop
``Nonlinear Physics. Theory and Experiment. IV'' (Gallipoli, Lecce,
Italy, 2006) for kind hospitality and the Russian Foundation for
Basic Research for Travel Grant No.~06-02-26768.

\end{document}